\documentclass{article}

\usepackage[utf8]{inputenc}
\usepackage[obeyspaces,spaces,hyphens]{url}
\usepackage{paralist}
\usepackage{natbib}
\def\citeN{\citet}
\def\citeANP{\citeauthor}
\def\cite{\citep}

\newcommand{\eg}{\textit{e.g.}}

\newcommand{\etc}{\textit{etc.}}
\newcommand{\ie}{\textit{i.e.}}

\newcommand{\vs}{\textit{vs.}}
\newcommand{\vv}{\textit{vice versa}}

\newcounter{usNum}
\newenvironment{userstories}{\begin{description}}{\end{description}}
\newcommand{\userstory}[3]{\stepcounter{usNum}%
  \expandafter\xdef\csname USN#1\endcsname{\arabic{usNum}}%
  \expandafter\xdef\csname USP#1\endcsname{~#2}%
  \expandafter\xdef\csname US#1\endcsname{\csname USN#1\endcsname~---~#2}%
  \expandafter\gdef\csname USI#1\endcsname{\textit{\csname US#1\endcsname\/}}%
  \expandafter\gdef\csname USA#1\endcsname{\item[\emph{User story
    \csname USN#1\endcsname\ --- #2\/}] #3}%
  \expandafter\gdef\csname UST#1\endcsname{\textit{\csname USP#1\endcsname~(\csname USN#1\endcsname)\/}}%
}

\begin{document}

\title{The Value of User-Visible Internet Cryptography}
            
\author{PHILLIP J. BROOKE\footnote{School of Computing, Teesside
University, Middlesbrough, TS1~3BA.  \texttt{pjb@scm.tees.ac.uk}}\\Teesside University
  \and RICHARD F. PAIGE\footnote{Department of Computer Science,
University of York, YO10~5GH.  \texttt{richard.paige@york.ac.uk}}\\University of York
}

\maketitle

\begin{abstract} 
  Cryptographic mechanisms are used in a wide range of applications,
  including email clients, web browsers, document and
  asset management systems, where typical users are not cryptography
  experts.  A number of empirical studies have
  demonstrated that explicit, user-visible cryptographic mechanisms
  are not widely used by non-expert users, and as a result arguments
  have been made that cryptographic mechanisms need to be better
  hidden or embedded in end-user processes and tools.  Other
  mechanisms, such as HTTPS, have cryptography built-in and only
  become visible to the user when a dialogue appears due to a
  (potential) problem.  This paper surveys deployed and potential
  technologies in use, examines the social and legal context of broad
  classes of users, and from there, assesses the value and issues for
  those users.



\end{abstract}

\section{Introduction}

Cryptography mechanisms are embedded in a range of software
applications, including Internet banking and online shopping. These cryptographic
mechanisms are, in some cases, entirely hidden from end-users; other mechanisms 
require the users to interact with them directly, and we call these
\emph{user-visible applications of cryptography}.  These mechanisms may involve users
entering passwords or passphrases for secret keys; 
other examples include dialogues related to resolving
problematic SSL certificates on web sites. 
The types of application that we are concerned with include email, web
browsing, e-commerce and document management systems; these applications are
used widely, particularly by non-IT expert users. In
all cases, the interactions that an end-user has with cryptographic mechanisms and
applications take place in a social and legal context.  This context includes the user's objective, 
such as buying a book from an online store or communicating with friends or colleagues.  

This paper surveys and analyses the use and impact of these
cryptographic mechanisms and techniques for general users.
We highlight  common methods and techniques, along with  problems in their deployment.
We start by describing the legal context of these non-expert end-users, examine security usability issues, then outline the technology at hand.  
We then use scenario-based analysis to structure a thematic survey of the applications of user-visible cryptography.  

We draw the themes together and discuss issues such as trust, deployment, endpoint security and their overall effect on the use of these cryptographic mechanisms, given the users' context.
We will conclude that in general, other mitigations are important in these user interactions, and 
that substantial automation is appropriate where cryptography is required.
We also argue that there are some limited cases where user-visible applications of cryptography has 
significant value
for typical end-users.

\userstory{wsocial}{Browsing a social website}{Alice reads and sometimes posts on a
  social website, \eg, FaceBook or web forums.  We suggest that the overall
  risk here is low: antisocial behaviour and account hijacking are the
  main risks, but the assets concerned are limited, at least from Alice's
  perspective.  A greater risk might be posed by Alice
  posting something she later regrets.}
\userstory{wecommerce}{Buy goods via a website}{Alice wants to buy something from an
  e-commerce site.  She will necessarily use her credit card or a
  service like Paypal.  Either way, at some point, she has to pay
  money in the expectation that the purchase is delivered as
  specified.  The risks are high here: phishing, website spoofing and non-delivery of goods
  are the canonical examples, along with theft of payment and other
  details from the recipient site.}
\userstory{wbanking}{Online banking}{Alice views account details and pays bills using
  her bank's online service.  The risks here are as in the previous
  story: bank details have an obvious value to criminals.}
  \userstory{esocial}{Social email}{Alice wants to email her relative or friend,
    say, Bob.  The overall risk is low: the main asset is the email,
    and it is unlikely to be particularly valuable although
    potentially embarrassing.  From Bob's perspective, someone
    pretending to be Alice is a very low risk.}
\userstory{esensitive}{Sensitive discussion by email}{Suppose Alice and Bob work
  together and need to discuss a serious problem with a particular
  task.  Email is one possible medium.  The risks revolve around confidentiality.}
\userstory{econtract}{Agree a contract by email}{Alice agrees by email to undertake
  some work for for a small business.  The main risk here concerns
  non-repudiation by the business or \vv.  Thus it is not so much an
  issue of making the contract but one of evidencing that the contract
  has been properly made, \ie, that the elements of consideration,
  intention, offer and acceptance are all present.}
\userstory{supdate}{Install or upgrade software}{Alice installs some software from the
  Internet.  How can she be sure that it is free of malware and from
  the correct publisher?}
  \userstory{finternal}{Internal application process}{An applicant, with the assistance of his
  supervisors, completes part of an application form.  Two different
  department heads need to sign off various resources and indicate
  their support, as well as obtaining certification from a finance
  department clerk.  This documentation is then forwarded for a final
  decision to be made.  

  The risks are a little more subtle than some
  of our earlier user stories.  If everyone is cooperating and
  trustworthy, there is no problem.  However, some people do attempt
  to defeat the checks-and-balances in such schemes: we discuss this
  further below.}
\userstory{fsign}{Signing a form}{A variation on the agreement of a contract: a
  publisher requests that Alice signs an agreement, \eg, our
  motivating example here is a transfer of copyright form.  As in the
  earlier example, this is a low risk example: the problem is to be
  able to provide evidence if the agreement was subsequently
  challenged.}
\userstory{data}{Data confidentiality}{Alice has some data on a laptop computer 
  that is the subject of  data protection obligations.  Laptops can be
  lost or stolen relatively easily, thus the risk is medium or high.}

\subsection{Structure}

First, we outline the legal context for our assessment of cryptography in
section~\ref{sec:legal-context}: this work is initially from an English and Welsh
perspective, but acknowledges the cross-border aspects of electronic
transactions.  We continue by surveying existing related work
concerning usability and security in
section~\ref{sec:user}.

Section~\ref{sec:tech} briefly introduces the common, underlying
technology that concerns this work.  We address emerging approaches in later discussion.
Section~\ref{sec:method} sets out the scope of our survey.  It describes the type of user we are concerned about (a ``general Internet user'' stereotype) and introduces ten user stories that we use to motivate our later discussions.   Section~\ref{sec:categories} identifies three categories of user-visible applications of cryptography.

Sections~\ref{sec:web}--\ref{sec:dfe}  examine the application of cryptography, grouping the user stories together thematically.
We consolidate and expand the discussions in sections~\ref{sec:web}--\ref{sec:dfe} and make some overarching comments in section~\ref{sec:discussion} before
concluding in section~\ref{sec:conclusions}.

\section{Legal context}\label{sec:legal-context}

This work is based primarily in a English and Welsh legal context.  However, the general
observations should be sound in similar jurisdictions, particularly derived legal systems
such as Canada and Australia. Moreover, the observations related to issues of data
protection and transmission (which impact on several scenarios) are also applicable to
European jurisdictions subject to EU~directives.

We structure this part of the discussion into two broad areas: contracts and signatures (integrity matters), and confidentiality and privacy.

\subsection{Contracts and signatures}\label{sec:contractsSignatures}

The formation of contracts
requires agreement, and that agreement is ideally recorded.
However, a contract can be made verbally as well as by hand-written
signature, but the burden of demonstrating a verbal agreement is greater.  
Contracts may impose confidentiality and similar
requirements on one or more parties.  

More relevant to computing, e-commerce requires that the making of
contracts is mediated by computers.  Thus a simple email indicating
agreement, or completing an online form by clicking ``accept'' can be
sufficient to form a contract.  This brings us to the use of
``signatures'', where examples include hand-written signatures, stamps
or images of a company officer's signature, email signatures that are
automatically appended, typed signatures, and so on, all the way to
cryptographic digital signatures.  Thus we see that ``signature'' is
a rather ambiguous term:  Gutmann's tutorial slides include greater
detail in this
area~\cite{mGutmann1}.
Additionally, \citeN{mMason1} gives a summary of some forms of electronic
signature, and comments ``the person relying on the signature (such as
where you say you did not sign a cheque, and the bank has paid money
out of your account on a cheque) must prove it was your signature
where you dispute it was not your signature. This is the same for
electronic signatures, although the vendors selling digital signatures
try to reverse this rule.''  
A thorough coverage of the legal issues surrounding electronic signatures is in Mason's book~\cite{Mason12}.

Some legislation explicitly addresses the recognition of electronic
signatures, such as the Electronic Communications Act
2000~\cite{ECA2000}.  
A result of this is that a wide range of statements can be legally considered an
``electronic signature''.  For example, Monitor, a regulatory body for
part of the UK's National Health Service, interprets this to mean that 
\begin{quote}
  ``the following are all examples of an electronic signature
  \begin{itemize}
  \item Typed name \item E-mail address \item Scanned image of a signature \item
    Automatic e-mail
    signature''~\cite{Monitor08}
  \end{itemize}
\end{quote}
This point applies to other media, such as faxes.
Chapter 6 of~\citeN{Mason12} provides a detailed analysis of the form of electronic signatures and comments on cases illustrating the variability of legal decisions.  
Mason also quotes the Law Commission writing on `Electronic Commerce', including ``[\ldots] the validity of a signature depends on its satisfying the function of a signature, not on its being a form of signature recognised by the law'' and ``Even if a click is less secure than a manuscript signature, reliability is not essential to validity.''  
This illustrates a distinction between the validity of a signature (essentially its acceptability) and the reliability of the method or form of the signature.

The later Electronic Signature Regulations 2002~\cite{ESR2002}
introduce the notion of an ``advanced electronic signature [which
means an electronic signature]
\begin{quote}
  \begin{enumerate}\def\labelenumi{{\rm (}\alph{enumi}\/{\rm )}} 
  \item which is uniquely linked to the signatory, \item which is
    capable of identifying the signatory, \item which is created using
    means that the signatory can maintain under his sole control, and
    \item which is linked to the data to which it relates in such a
    manner that any subsequent change of the data is detectable''
  \end{enumerate}
\end{quote}
where an electronic signature itself ``means data in electronic form
which are attached to or logically associated with other electronic
data and which serve as a method of authentication''.  Additionally, \emph{qualified
certificates} are introduced which have additional liability provisions.
These Regulations follow
the European Directive 1999/93/EC on a Community framework for
electronic signatures~\cite{mEUESa,mEUESb}
 ``addresses three forms of electronic
signatures: Basic electronic signature [\ldots] Advanced electronic
signature [\ldots] ``Qualified electronic signature'' [\ldots]'' and
these are criticised in~\citeN{Krawczyk10}.  
\citeN{Mason12}~comments that the European Commission ``may make further efforts to encourage the take-up of digital signatures, in the face of overwhelming evidence that nobody seems to want to use them, unless they are forced to do so.''
A more general coverage of the evolution of documents and the use of technology, including cryptography, is given by~\citeN{Blanchette12}.  Although sometimes from a French

Information from computer records themselves also has value.  The
Civil Evidence Act 1995~\cite{CEA1995} specifically places weight on
the evidential value of a computer record rather than the
admissibility of the record itself.  Thus records of businesses and
public authorities can be relatively easily used as evidence.
A similar provision exists in the US court system in the form of  Rule 803~\cite{mUSevidence}.

There is recognition of the need for reliability in the processes surrounding computer systems and evidence.  For example, BS~10008~\cite{BS10008} and the associated BIPs~\cite{Shipman08,Shipman+08,Howes08} provide substantial guidance; supplementary material includes a workbook to assist audit.

Note that other than advanced electronic signatures and some references in BS~10008, nothing above
explicitly requires any form of cryptography.  Indeed, this, at least
so far, poses no problem for the making of contracts in this context.

\subsection{Confidentiality and privacy}

A major issue for information systems concerns data protection
legislation, primarily, the Data Protection Act 1998~\cite{DPA1998},
an enactment of the 1995 European Union Data Protection Directive. 
This imposes obligations; 
for example, principle 7 states ``Appropriate technical and organisational measures shall be taken against unauthorised or unlawful processing of personal data and against accidental loss or destruction of, or damage to, personal data.'' 
The definition of “appropriate” is, of course, subject to each individual case.
Formal notions of “data controller” (a person or persons responsible for the processing of data) and “data processor” (for outsourcing of processing) are given in this Act.
Substantial guidance exists, along with a range of standards such as the ISO 27000 series.
Besides regulatory requirements, information usually has value to both individuals and
businesses, regardless of the presence or absence of personal data.
All this needs protecting in the traditional senses of
confidentiality, integrity and availability.

Classic examples involve sensitive medical records, bank account and credit card credentials.  
Within the UK, the Information Commissioner's Office is responsible for enforcement.  
Some remedies are also available for data subjects (such as demanding the correction of erroneous records).  
We remark that public reports of legal action following security
lapses are unusual, with Sony's recent security problems being an
exception~\cite{mSony}.  However, such lapses tend to be either
failure of access control or loss of devices or media with plaintext
data.  We discuss this further in section~\ref{sec:discussion}.

Privacy is related to, but not synonymous with confidentiality.  
In this survey, we do not need to consider these difference further with the exception of noting the recent regulations regarding cookies~\cite{mICO2} due to European Directive 2009/136/EC.
Compliance with these regulations is interesting due to the contrast with consent (partially discussed above in relation to clicking “accept”): for example, “Implied consent is a valid form of consent”.
Additionally, there are  broader matters of (mis)use of web technologies (including cookies) in malware and surveillance. 

\section{Usability and security}\label{sec:user}

Previous work assessing the effectiveness and value of cryptography
has examined usability as well as PKI issues.  These
areas dominate this paper, so we discuss them here.  
We also introduce further literature where relevant in the sequel.

A classic paper in the usability field is~\citeN{Whitten+05} which
concerns the ability of users to use PGP~5.0: ``Our 12 test participants were generally educated and experienced at using email, yet only one-third of them were able to use PGP~5.0 to correctly sign and encrypt an email message when given 90 minutes in which to do so''.  More generally, Furnell
and others have investigated the usability of end-user software at
length, and find continuing problems with
interfaces~\cite{Furnell+06,Furnell07,Ibrahim+10,Sweikata+09,Cranor+05,Gutmann+05}.
\citeN{Ho+10}~examined the setup of home wireless networks, and found
that “users did not understand the difference between access control
lists and encryption, and that devices fail to properly notify users
of weak security configuration choices”.  They proposed a configuration
wizard to partially mitigate some of these problems.  \citeN{Zurko+96}
introduced the term “user-centered security” and discussed the
application of usability testing to secure systems.  Some attention
has also been paid to the education of users in the use of security-related
software~\cite{Reid+05}.  

Others comment on the software itself.  \citeN{Kapadia07} remarks ``I found that
[OpenPGP applications] were unusable with nontechnical correspondents
because it required them to install additional software'', which
relates to some of our remarks on systems such as IronPort and
Hushmail in section~\ref{sec:gateways}.  We used a similar approach of
server-side cryptography in support of document
security~\cite{Brooke+09b}.  

Other work assesses what the users
understand about security concerns: \citeN{Gross+07} interviewed
twelve users with differing roles to answer “What do users know about
security and threats?”; “How do users manage their security concerns?”
and “Who do users believe is responsible for security, and how do they
perceive their role in security?”, noting that “entire organizations
can be brought down by security failures”.  Later work suggests that
users do differentiate between security (and privacy) concerns and
more general computer problems (such as hardware
failure)~\cite{Gross+07b}.

Previous work has also examined PKIs and questioned their
effectiveness and usability~\cite{Gutmann03,Straub+04}.  Moreover the
\emph{need} for PKIs, electronic signatures, \etc\ is not clear in
practice~\cite{Bileta11a} (and our earlier comments in section~\ref{sec:contractsSignatures}).  Alternatives involve opportunistic
encryption~\cite{Garfinkel03}, key continuity
management~\cite{Gutmann04b,Garfinkel+05}, identity-based
encryption~\cite{Shamir85,Martin06} and email-based identification and
authentication (EBIA)~\cite{Garfinkel03b}.  However, we are not
concerned with some other security properties, such as anonymity in
systems such as Mixminion~\cite{Mathewson+04}. 

More broadly, notions of
\textit{return on security investment}~\cite{mNIST05} attempt to capture
the return on investment in security processes, policies and infrastructure,
though this focuses on capital investment rather than value
delivered to end-users. 
An interesting variation is due to~\citeN{Herley09}, who argues that
users' rejection of much conventional security advice (for example,
ignoring SSL certificate warnings) is rational.  This is on the basis
of out-of-date advice and false positive warnings against the cost (to
the end-user) of acting on this information.  Herley examines password
rules, phishing site identification and SSL certificate warnings and
comments “the burden [to the end-user] ends up being larger than that
caused by the ill it addresses”.  Similarly, \citeN{Bohme+11} argue
that human attention is a scare resource.  They too make the point
that user inattention can be rational, and produce a simple game model
to illustrate typical options for users.  A possible way to reduce the
demand on attention is the use of social navigation, as suggested by
\citeN{Goecks+09}.  They present prototype tools which describe other
users' security decisions (\eg, for cookies and firewalls), although
it proved less useful for more complex or ambiguous decisions.

\section{Underlying technology}\label{sec:tech}

Cryptographic technologies typically address confidentiality and
integrity issues.  The underlying mathematical concepts of these
technologies are the same: both employ a range of asymmetric and
symmetric algorithms (\eg, RSA and AES respectively).  Typical operations include
\begin{itemize}
\item key generation, both for long-lived public/private keys as well as transient session keys;
\item encryption and decryption (confidentiality);
\item signing and verification (integrity); and
\item hashing (\eg, as part of signing, or deriving a key from a password or passphrase).
\end{itemize}
We do
not dwell on the mathematical approaches (an appropriate starting
point is~\cite{Schneier96}), but instead on how they are encapsulated
into the applications and made visible to the user.  
Later, in section~\ref{sec:interop}, we see that this encapsulation is not trivial;
for example, different software can interpret normalisation of messages in different ways resulting in false bad verification of signatures.

As well as understanding the basic capabilities and scenarios of interest
for non-expert end-users (section~\ref{sec:users}), we also must clarify the technical context in
which they work. We briefly summarise several major groups of cryptographic software;
our end-users will likely use one or more of them either
explicitly or implicitly.
\begin{description}\def\makelabel#1{\it #1}
\item[CMS] or Cryptographic Message Syntax, based on PKCS\#7, is described by
RFC5652~\cite{RFC5652} and describes a message format for
cryptographic messages.  It is usually used alongside an X.509
public key infrastructure.  The best example of CMS is its use in
S/MIME email messages.
\item[X.509]  itself defined in RFC5280~\cite{RFC5280},
provides the most common format for  public key  infrastructure (PKI) data for the
Internet.  This usually leads directly to the certificate authority
trust/validity model.  
Certificate revocation lists are also supported in X.509; however, the Online Certification Status Protocol (OCSP)~\cite{RFC2560} perhaps provides an alternative giving more timely updates.
\item[SSL/TLS]   Significantly for our example users, SSL/TLS is widely
  deployed on websites and mail servers.  (Although not identical, we use SSL and TLS as synonyms in this work.)  In this role, it is a
  near-ubiquitous protocol with  native support in common web and
  mail clients.  
  Public keys (as X.509 certificates) are obtained in the initial SSL/TLS negotiation.  
  The relying party then needs to verify that the presented certificate is signed by a trusted root certificate, possibly via intermediaries. 
  We return to this issue in section~\ref{sec:trustRoots}, including comments on alternative approaches.
\item[OpenPGP] defined in RFC4880~\cite{RFC4880}, is an alternative to
  S/MIME for email messages as well as for general file encryption and
  signing, based on Zimmermann's PGP.  X.509 
  certificates are not used in OpenPGP; instead a web of trust is
  usually used instead.  The web of trust is not the only option:  single
  and multiple key validation models are supported.  
  Both PGP and GnuPG support this standard and are broadly interoperable.
\end{description}

Other interesting technologies timestamping services, key servers (for
OpenPGP keys), and other means of obtaining up-to-date keys, such as
integration into Active Directory and LDAP.  The Simple Public Key
Infrastructure
(SPKI)~\cite{RFC2692,RFC2693,mSPKI},
described in experimental Internet RFCs, concerns a more local naming
scheme.  We will return to some key management issues later in the
discussion.  More user-friendly approaches include Hushmail and
similar services (discussed in section~\ref{sec:gateways}).

\section{Users and user-visible applications of cryptography}\label{sec:method}

We will define what we mean by \emph{user-visible applications of
  cryptography} in section~\ref{sec:categories},  and first describe the type of users we are
concerned with.  

\subsection{Users of interest and scenarios}\label{sec:users} 

The typical users of interest are
\begin{enumerate}
\item  domestic users with tasks such as social email, online shopping and
  e-banking;
\item  office workers, using software such as office productivity
  applications, undertaking sensitive discussions by email, or working
  with sensitive data such as personal data;
\item supervisors and managers interacting with other staff,
  authorising, approving and auditing business processes;
\item non-IT-specialist users installing or upgrading software, \eg,
  operating system updates, plugins such as Adobe Flash and entertainment software.
\end{enumerate}

A scenario-based approach~\cite{Carroll+98,Rosson+02} allows us to
structure the analysis by end-user concerns. From an analysis of the
literature and incidental observations of end-users we identified a
set of ten typical scenarios where cryptography plays a role:
\begin{quote}
  \begin{tabbing}
    \USwsocial \\
    \USwecommerce\\
    \USwbanking\\
    \USesocial\\
    \USesensitive\\
    \USecontract\\
    \USsupdate\\
    \USfinternal\\
    \USfsign\\
    \USdata\\
  \end{tabbing}
\end{quote}
Browsing social websites was common to most computer users, with
Facebook particularly prevalent.  Most users had experience of
ordering from the Internet, such as Amazon, and using the Internet for
banking.  Social email is perhaps less common than previously (we
speculate that social sites such as Facebook account for this;
however, this was not investigated further), although all used email
as part of their work.  Those in sensitive areas (healthcare, criminal
justice) often engaged in discussions of cases by email.  Few had
agreed formal contracts by email, but negotiations that had an impact
on subsequent contracts were commonly mediated by email.  Nearly all
users had installed software, often games or plugins as well as
downloading applications to mobile devices (\eg, iPhones, Android
devices).  The larger organisations had formal processes that involved
rigid workflow processes as well as requirements to ``sign'' forms
in some way.  The final user story, dealing with confidentiality of
data, concerned those users working with “personal data”.

The ten scenarios are a representative set to allow
us to break down user interactions with cryptography.  We do not claim
they are complete; there are other specialised cases that we do not
attempt to address.  Instead, we are concerned with a ``general
Internet'' stereotypical user without specialist skills or needs; 
we do not address scenarios such as the use of ATM cards or RFID-based and similar access control systems.
Later sections group these user stories thematically;
subsequently, we consolidate the points in
section~\ref{sec:discussion}.  

Before we can analyse these scenarios, 
we must say more about our assumptions relating to our users
and their environment.
As we have suggested already, we do not address relatively small,
specialised user groups with very high security demands.  These
specialised populations can reasonably be expected to undertake
appropriate training and be supplied with suitable equipment for their
tasks.  Instead, we are interested in day-to-day use of computers.

A common assumption to all these users is that they have basic
computer skills, \eg, word processing and email, but they are not IT
specialists and have no need (nor interest, often) to be IT
specialists.

Our analysis required us to make assessments of risk.  We followed the
common method of identifying the likelihood as low, medium or high,
and the impact as low, medium or high.  A typical approach then
assesses the overall risk as low, medium or high from the likelihood
and impact.  In the sequel we discuss the risks identified, starting
with the highest.

\subsection{User-visible applications of cryptography}\label{sec:categories}

There are three categories of user-visible applications of cryptography that we concern ourselves with here.

\begin{enumerate}
\item\label{item:direct} The most obvious user-visible application of cryptography is the
  direct, elective invocation of a cryptographic tool, \eg,~PGP or
  GnuPG.

\item\label{item:indirect}  Indirect but still explicit, elective use of cryptography
  involves examples such as
  \begin{itemize}
  \item asking an S/MIME email client (\eg,~MS~Outlook) to encrypt or
    sign an email;
  \item encrypting or signing a document in an office application
    (\eg,~MS~Office, LibreOffice); or
  \item selecting encryption in a ZIP archive application
    (\eg,~7-Zip).
  \end{itemize}
  Sometimes this is a simple as ticking a box to select encryption and
  giving a password which is subsequently used (in some form) as a key
  to a symmetric algorithm.  Others, such as signing office documents,
  requires at least a user certificate for an asymmetric algorithm or
  a full PKI.

\item  Much cryptography occurs in the background.  Web browsers and email
  clients can automatically use SSL, discussed further in
  Sections~\ref{sec:web} and ~\ref{sec:email}.  This is implicit and
  should be unobservable by the user until there is a problem, such as
  an out-of-date or otherwise invalid certificate causes the client
  software to warn the user.
\end{enumerate}

The examples above in categories~\ref{item:direct} and~\ref{item:indirect} usually affect the recipient.  A signed document
might not require any special interaction, yet the client software may
report the state of the signature, possibly raising dialogues or
showing warnings.  In other cases, the recipient may be completely
unaware of the signature (\eg,~an office document with an embedded
signature, or a multipart signed email) or conversely, the document
may be unreadable without using specialist software (such as
ASCII-armoured signed emails).  

Encrypted emails and ZIP archives
nearly always require a direct interaction to give the relevant key,
usually in the form of a password or passphrase.  For an email, the
relevant private key may already be accessible for automatic
decryption, as in some configurations of MS~Outlook.

Similarly, the first two categories may require explicit key management on part of the users.

In this survey, we concern ourselves with the examples above where the user becomes
aware of the presence of some problem or issue in the underlying structure.  
Importantly, the user does not  necessarily have to  relate this to a cryptographic system at all; consider \citeN{Ho+10}'s comments on users not understanding the difference between different concepts.  

Although we will discuss some issues of endpoint security, mostly in section~\ref{sec:endpoint}, this is in relation to the overall risk for different scenarios.  Thus we do not discuss the use of passwords and other authenticators beyond that.

\section{Web}\label{sec:web}
We now examine the ten user stories, grouped thematically.
We start with three typical web-based scenarios.

\begin{userstories}
\USAwsocial
\USAwecommerce
\USAwbanking
\end{userstories}
Although relatively obvious, we can find illustrations of these user
stories in Anderson's text~\citeyear{Anderson08} in sections~23.3.3,
23.3 and~1.3 respectively, and additionally for the latter two user
stories in~\citeN{Cronin97}.  Evidence of interest in social
networking more broadly can be seen in~\citeN{mSocialnets09}.

Secure web connections via the HTTPS protocol are relevant to these
three user stories.  In each case, Alice will have to point
her web browser to the correct URL: this URL might have been
bookmarked from a previous visit, found via a search engine or typed
in, perhaps from an advert in a newspaper, or from memory.

Before we consider HTTPS directly, let us address the risks.
The main risk is the compromise of login credentials: these
credentials are useful to attackers for harassment/nuisance via
social media, theft from online banking or misuse of credit card
details. Compromised credentials can then be used to call into
question the integrity of any transaction involving those credentials
or to present the possibility of compromised credentials for
``plausible deniability''.  Additionally, the re-use of passwords,
even on ostensibly low-security websites clearly permits further
exploitation of credentials: ``a substantial number of the randomly verified email accounts revealed that 75 percent of the users rely on the same password to access both their social networking and email accounts''~\cite{mBitDefender1}.

At some point, payment details are required.  The web browser is
redirected to a ``secure page'' accessed via HTTPS if the entire site
is not already HTTPS based.  At this point, we encounter our first
problem.  The reliance on certificate authorities for X.509
certificates to bootstrap what is essentially a trust relationship has
been highlighted previously~\cite{Perlman99} and was brought sharply
into focus with the Comodo compromise in 2011~\cite{mComodo1} and the more recent issues with DigiNotar~\cite{mLWN1}.  A
secondary issue to the Comodo and similar compromises concerns the limited use of
CRLs and OCSP by clients to revoke bad certificates.
\citeN{mMozilla}~ reports that the offending certificates were quickly
revoked using both the CRL and OCSP mechanisms.  We see other examples
of this later.  Trust in the computers concerned is a deeper
problem~\cite{Parno+10}.

Common advice given to users for e-commerce transactions typically
includes “Check that the padlock sign is shown on your browser and
that the URL includes \verb|https|.”  Regardless, users still find it difficult to assess whether or not “a connection [to a web site] is secure”~\cite{Friedman+02}. Complications include extended
validation and more sophisticated phishing attacks~\cite{Jackson+07}.  \citeN{Kirlappos+12} argue that trust seals are ineffective, and conclude that “automatic verification of authenticity” is required.
Rapidly changing browser environments are also likely to confuse users; for example, Mozilla Firefox has changed its indication of secure connections several times~\cite{mSchultze1}.
What Alice really needs is sufficient evidence that her web client
is connected to the correct server and that the connection to that
server is encrypted.

Observation of some user populations at our institutions (in our
cases, academics and students) demonstrates that the security afforded
through CAs is brittle at best.  Warning dialogues are often
disregarded~\cite{Likarish+08}: we have effectively trained our users
to ignore the warnings because they have to workaround problems.  One
of the ICT departments at the authors' institutions included
instructions to set up a wireless connection which explicitly
directed the user to accept an invalid certificate because of the
server's setup.

The difficulty in assuring that the client has
connected to the correct server is one factor that enables phishing.
In one sense, this is an artefact of a global naming scheme (the DNS)
and we see that SPKI suggests local naming schemes in closed groups.
But this poses difficulties for, say, the banking scenario.

A moderately na\"ive solution for online banking would be for a bank to
tell users the fingerprint of the correct certificate: but we do not
believe that any but the most security-conscious user would actually
check this.  Essentially, the computer is a tool and fine management
of it is simply not a conscious matter for the user.  Hence our focus on \emph{user-visible} applications of cryptography.

The issue of root certificates aside, the actual usability is
relatively good: we do not see people having great difficulties making
e-commerce purchases.
We return to this in our discussion in section~\ref{sec:discussion}.

\section{Email}\label{sec:email}
Our next set of scenarios relates to use of email, at different levels of
sophistication and hence, with different requirements for use of
cryptography.

\begin{userstories}
\USAesocial
\USAesensitive
\USAecontract
\end{userstories}
An example of sensitive email is given in \citeN{Gaw+06}.  Movement of
email services into the ``cloud'' is advancing, with outsourcing to
Google and Hotmail in evidence, and along with suggestions for the US Federal Government~\cite{mSaaSemail}.

Email is, for many users, an effective communication medium, although
the prevalence of both spam, which we do not directly address, and
large volumes of legitimate email can degrade this.  The essential
risks here are twofold: one is the loss of confidentiality, the second
risk concerns spoofing or modification  (integrity) and non-repudiation.

This is a good example for opportunistic encryption~\cite{Garfinkel03}.  A mail user
agent or mail submission agent connecting to a server may use
SSL/TLS to encrypt the conversation with the server.  This has the
same problems as for web servers, \ie, how does the user know that
they have connected to the correct server?  But differently from the
HTTPS example, mail servers are arguably harder to spoof.  Two
major classes of mail server are those within a particular business
and those for the user's ISP.  In both cases, we should have a good
level of confidence that the relevant part of the DNS is correct, at least from the
client's perspective, and that regardless of the
certificate, we have connected to the correct server. 
Some users may be in a closed or partially restricted environment (\eg, heathcare) further reducing the incidence of problems.
However, this
observation leads us to a further point: within a particular business,
how many users are likely to be actively sniffing the network?

We develop this point further.  Older, hub or broadcast-type networks
are very easy to monitor for other users' traffic.  Newer switched
wired networks are harder to monitor although some switches are
believed to degrade to operate as hubs.  Wireless connections are an
instance of broadcast networks, which are potentially easier to
monitor unless encrypted, say, WPA2.  Since it is relatively cheap
and easy to arrange for a mail server to offer SSL/TLS connections, it
is proportionate to do so and thus not worry about any possible
sniffing by insiders or those with access to the network.

Mobile users provide a complication.  The argument above does not
apply to a user temporarily visiting another organisation or using a
hotspot as they cannot rely on the infrastructure to the same degree
(for example, there is more delegation of DNS).  This is no worse than
the general HTTPS case.

Thus for most users, they can assume that their ISP or business mail server
does receive their email, and opportunistic encryption using SSL/TLS
defeats any local sniffing.  However, if the ISP or local mail server
is not trusted, the user may be reluctant to trust this encryption of
the connection.  Further, this is  only transport encryption, not storage
encryption.  The email must be stored on the mail server, even if only
transiently, as email must be stored temporarily on each server that
handles it.  Certainly in the case of a business mail server, there
is a significant broader problem: if the users cannot trust their own
servers, then what else is wrong with the infrastructure?

\subsection{S/MIME and OpenPGP email}

This leads us to consider S/MIME and OpenPGP for emails.
Capable users might choose to generate key pairs and use one of these
cryptosystems to ensure confidentiality of their messages.  However,
these are, by observation, a tiny minority of the population as a
whole.  One barrier to adoption of this approach for secrecy is the
need for the recipient to have a public key; this results in multiple
attempts to create public keys on demand, \eg, identity-based
encryption.  We discuss these and similar 
approaches such as Hushmail in section~\ref{sec:gateways}.

Moreover, within a particular organisation ---with an assumption of a
trusted infrastructure--- the emails are already safe due to
opportunistic encryption, other than at the endpoints.  These
endpoints are the sender's and receiver's computers.  Here, we can
remark that some user's security hygiene is negligible, \eg,
our remarks about screenlocks on page~\pageref{pg:screenlocks}.  It
is, of course, notable that users' desktop machines are a major entry
point of malware, via the web or USB sticks.  For example
\citeN[slides 108--109]{McQueen10} reported that 20\% of users inserted a
thumb drive found in a public place into their computer.  In our discussion
(section~\ref{sec:discussion}) we further comment on endpoint security.

Returning to the point of opportunistic encryption, we note that
discovery of the correct settings can be challenging.  We speculate
that increased outsourcing of email services in large organisations
may be to blame.  Some software, such as Apple's Mail, seems
remarkably robust.  Mozilla's Thunderbird needed much help to connect
to the student mail system at one of the authors' institution.

Additional aggravations concern the use of passwords and passphrases
used for securing cryptographic keys.  For example, some systems do
not require a password after importing a PKCS12 file: the private key is
accessible on demand.  Thus someone with access to that desktop
machine can read any email, even if it is encrypted to that particular
key.

Further, key management remains a major problem.  \citeN{Gutmann03}
reports that obtaining a key from a public CA ``takes a skilled
technical user between 30 minutes and 4 hours work''.  Little has
changed since then, and in any case, these certificates are ``low
value''.  Local CAs using the SPKI model can more easily issue
certificates for their own servers, and can ensure that centrally
provisioned machines have the relevant root certificate installed.
But external users do not benefit from this.

The problem goes on step further.  We have seen examples of users in
the public sector sending emails with S/MIME signatures.  “Good”, one
might think.  However, the certificate issuer is one of these local CAs: we can
decide to accept the issuing certificate in our mail client.  But some
software, such as \verb|gpgsm| takes the decision
that certificate revocation lists must be checked: this is correct in our view.  At this point, we
discover that the machine which serves the CRL is not accessible
outside of that organisation.  The value of the CRL, and thus the
certificate overall, is massively reduced.  Moreover, the particular
characteristics of the organisation in our example make it very
unlikely that unauthorised users would have access to even that organisation's buildings, let alone the computers within them.

Even if a user perseveres and obtains a key for use with their email
client, configuration and setup often remains challenging.  Dialogues
remain unintuitive for the most part.  In the course of other work, we
counted 8--9 steps
to import an S/MIME certificate from a PKCS12 file, depending on email client.

So we assert that there is no real value in signed email except in the
case where users have both good reason to
fear spoofing or modification or their messages, \emph{and} when they
have had opportunity to confirm, ideally face-to-face, that the
cryptographic certificates are correct.

\subsection{Interoperability and robustness}\label{sec:interop}

Even if we addressed the issues above, interoperability is poor in
contrast to general use of web browsers with HTTPS.
We examined a range of email clients, using versions current in early 2011, as listed in
Table~\ref{tab:clients}.
\begin{table}
  \centering
  \begin{tabular}{|l|l|l|}
    \hline
    &S/MIME&OpenPGP\\
    \hline
    MS Outlook&native&\\
    Mozilla Thunderbird&native&Enigmail plugin\\
    Apple Mail&native&\\
    Alpine&native \& filters&various filters\\
    \hline
  \end{tabular}
\caption{Email clients examined\label{tab:clients}}
\end{table}
Although S/MIME is generally well-supported natively, OpenPGP often
requires plugins and these are not available for some MUAs, notably
MS~Outlook.  This means that communities of users need to agree on
the cryptosystem to be used; yet these communities are often not
well-defined and have porous boundaries.

We sent and received emails using either S/MIME or OpenPGP.  For
OpenPGP, we examined both “inline” and MIME/OpenPGP messages.  
Encrypted messages were uncomplicated and mostly worked.  On some
occasions, they were simply not recognised and were ignored by the
client: the common feature in theses cases is that the
\verb|multipart/encrypted| message was not the top-level MIME part.
However, such messages are entirely valid in terms of MIME and
arguably, could occur in practice when digests are sent.

Verification of clearsigned messages was much more brittle.  Again,
some clients required that the \verb|multipart/signed| message was the
top-level, or it would be ignored and not displayed.  Others had trouble
verifying messages they had sent themselves!   Clearsigning is strongly
preferred over opaque messages, as clearsigned messages are readable
by users who do not have software capable of verifying the signature.

When we find that some mail servers also rewrite MIME messages causing
clearsigned messages to fail to verify, we conclude that the
technology remains too brittle and interoperability is relatively
weak.  This is disappointing after so many years.  The problems are
well-known, including suitable treatment of whitespace, line-endings
and character sets (indeed, we had to address the same canonicalisation
process when working with XMdoc~\cite{Brooke+09b}).  That email
systems remain so brittle in respect of clearsigned messages mitigates
against their use, as false negative verifications degrade the
usefulness of signing even further.  They lead to the same issue that
we encounter with web server certificates, where users are trained to
ignore the warning messages, if they actually check the signature at
all.  Indeed, we speculate that it would take other users a long time
to notice if we sent signed emails with a revoked key.

\subsection{Transparent solutions and gateways}\label{sec:gateways}

Sending an encrypted email requires that the recipient has a key to
decrypt it.  Both symmetric and asymmetric cryptosystems have
well-understood problems.

Identity-based cryptosystems are rooted in \citeN{Shamir85}'s work;
other work includes~\citeN{Martin06}.
Typically, the key generating centre is a trusted third party, and can
compromise the system.  This is not necessarily a problem, given that
some trust is required at some point.  \citeN{Boneh+01} provide an
example of an identity-based encryption system, and give several
useful properties such as restriction to dates and security
classifications, easy revocation and delegation of decryption keys.
\citeN{Cocks01} describes a scheme based on quadratic residues, and
comments that multiple authorities ``will be desirable''.  This point
is addressed by \citeN{Lee+04}, \citeN{Gentry03} and similar work, although the fine
details do not concern us at this point.  In general, we need to trust
some infrastructure, and simpler schemes have obvious single points of
failure and escrow.

A related approach is to make this as transparent to the end-user as
possible, particularly in terms of software requirements which we relate to
the earlier quote in section~\ref{sec:user} from \citeN{Kapadia07}).  We use
IronPort~\cite{mIronport1} and Hushmail~\cite{mHushmail2} as exemplars here.  Both can use a Java applet
so that decryption occurs on the client machine.  Additionally, both
offer an option for processing messages on the server machine via a secure web
session.  In this latter configuration, these services are not
significantly stronger than HTTPS as described above: this is
recognised in such services~\cite{mHushmail1,mHushmail2,mWired07}.
Some implementations send the email directly and only the decryption
key is escrowed, which has some positive impact.

A positive side effect is that policy engines such as IronPort can be
used to reduce the ``fat-fingering'' of emails by requiring that all
out-of-organisation emails are subject to policy enforcement (\eg,
encryption, or simply disallowing some outbound traffic).

Thus our point remains: for a typical user, what threats does this
mitigate?  The endpoints remain a problem: for example, some local
users of health service data receive messages via a secure email service of
a similar design as discussed above.  But the data is stored locally,
as plaintext.
If we combine local plaintext storage with a transparent approach and
implicit trust in the service provider, there seems to be little
security advantage over opportunistic encryption of email or a ``secure
web dropbox''.

Key continuity management (KCM)~\cite{Gutmann04b}, based on
imprinting~\cite{Stajano+99b} or trust-on-first-use (as in SSH), are further options: we implicitly
trust the first contact and only warn if credentials change
unexpectedly.  \citeN{Garfinkel+05} experimented with S/MIME, Outlook
Express and KCM, and concluded that ``KCM is more secure than today's
alternative to KCM: no cryptographic protection at all'' but also ``it
is not the panacea to the mail security problem for which we are
looking''.  
Related attempts include STEED~\cite{Koch+11}, which argues for end-to-end encryption and (similar to earlier points) trust-on-first-use.  STEED also includes further attempts to make key management easier: automatic key generation and key distribution via DNS.

\section{Software signing}

\begin{userstories}
\USAsupdate
\end{userstories}

The primary objective here is to ensure the integrity of the system
as a whole.  Once installed, operating systems typically receive
updates over their lifespan, for example~\citeN{mMS1} and \citeN{mDebian1}.  Application software is initially
installed and subsequently updated.  In all these cases, the intent is
to ensure that the “correct” software is installed or updated, in the
sense it should be “approved” or at least “certified” by someone
responsible.  The simplest case is that the original publisher or
developer has made the updates available, but there is a an obvious
competitive argument in favour of third parties making plugins,
updates, \etc, available.
Typical examples include drivers and updates on Microsoft Windows and
package signing in the Linux distributors, \eg, Debian's checking of
signatures via \verb|apt|~\cite{mDebian2}.

We make much use of ``scare'' quotes in the previous paragraph: the
exact purpose or value of the software can vary between stakeholders.
For example, some vendors may wish to restrict the platform so that
only software they approve is installed (perhaps for control of a
“marketplace”), or to limit potentially bad interactions of packages.

The risks are obvious: malware can masquerade as “genuine” software,
and thus we make the reasonable leap to cryptographically signing software.  We
observe that some security incidents in own institutions are due to attempts to
install software of relatively dubious origin.

From observation of users, we see two well-known
issues:
\begin{itemize}
\item As with the web and email examples, users
  disregard warnings because they obstruct the user's intention: to
  install some software.  

  Note that we do not concern ourselves with
  policy issues.  For example, some system administrators may wish to
  ensure that only particular patches are installed; involuntary
  upgrades may break other software.  Additionally, some patches are
  large, and may inconveniently use disproportionate amounts of
  bandwidth for roaming users.
\item Software signing uses public key cryptography: thus some public
  keys have to be trusted.  We have the usual root trust problem as
  described in section~\ref{sec:web}.  Indeed, this scenario can be
  viewed as a subset of the connecting-to-a-web-server scenarios.
\end{itemize}

\section{Form signing}\label{sec:formSigning}
This set of user stories is somewhat different, and relates to applications
in use in specific domains and industries --- particularly those with 
requirements for signing electronic documents.

\begin{userstories}
\USAfinternal
\USAfsign
\end{userstories}

We have previously examined the issue of distributed non-centralised
forms with requirements such as integrity and auditability
in~\citeN{Brooke+09b}.  One of the motivating scenarios there was our
current \USTfinternal\ user story.  A very specialised form of
signing (certification, in this instance) covers court documents, as
described in~\cite{Reiniger+10}.
Some services provide a web-centric approach, such as Adobe's EchoSign service~\cite{EchoSign}.

However, we now look at the broader process in the event of a
subsequent problem, and compare with the ``sign form'' scenario.

A non-computer approach for the latter scenario is for the publisher
to post the form to Alice, who signs it and posts it back.  A more
common method is to email a document, \eg, PDF, MS~Word, and ask for a
signed copy to be scanned then emailed or FAXed.  A final option (which
the authors have seen several times lately) is for the publisher to
offer an option of signing the PDF file using an X.509 certificate.
Again, this causes a dependence on certificate authorities as discussed earlier.

In common with the ``agree a contract by email'' user story, we note that an email
itself, even without any cryptographic measures, is likely to be
sufficient as evidence.  Similarly, due to legislation such as the Civil
Evidence Act 1995~\cite{CEA1995},  the document signing scenario is
relatively easy: we would assert that the computer system is functioning
correctly and the existence of the records would be sufficient.  One
party would have to actively dispute the validity of the assertion.
Of course, cross-border issues complicate this, but all agreements we
have seen include choice of law clauses, thus mitigating this issue.

Interestingly, we can raise difficulties with demonstrating that a
signatory has \emph{seen} and \emph{understood} the terms of an
agreement.  Click-through agreements are believed to be
enforceable~\cite{mMason1} although particular clauses may not be.

\subsection{Audit and integrity}

Complications concern the splicing of documents,
whether intended to subvert organisational controls, or simply to
expedite a process.  

We assert that people are essentially trusting.  Consider again the
\USTfinternal\ user story above; a more specific instantiation
is the approval of a course of training within an organisation (based
directly on a real system).  The process itself is relatively
involved, but a major problem in terms of audit and good governance
concerns the signing of these forms.  In a purely paper process, a
single document should be signed by all the parties (applicant,
supervisors, department heads and finance clerk).  But the difficulty
of obtaining all these signatures with an increasingly mobile
workforce often results in multiple signature pages being submitted
for a given document.  Worse, there is no guarantee that the
signatures are attached to the correct version of the document:
multipage documents can be easily spliced together.

The next step is to consider how these documents are handled when emails become
involved.  The committee that makes the final decision on these
documents now routinely sees a word-processed form, with some
signatures on printed pages and some printouts of emails from various
principals asserting their support of the application.  

In both the purely paper process and the process involving emails
there is trust that no one is actually trying to defeat the system by
presenting an application with putative signatures that are in some
sense false.  The emails are being sent within the same organisation,
using the same central mail service, and thus those relying on the
veracity of these emails and hand-written signatures trust the system
as a whole.  Essentially, we assert \emph{that there is no demand in
  typical domestic or business processes for cryptographic assurance of emails}.

Interestingly, this appears to be backed by the experiences with
qualified certificates (\eg, \cite{Krawczyk10}, referred to in
Section~\ref{sec:legal-context}): there is simply no real market for
them outside of very specialised demands.  This is likely to continue
while there is no statutory requirement to use an advanced electronic
signature or qualified certificate, since the existing legal framework
accepts the name on the email as being sufficient replacement for a
hand-written signature.  So a simple email is sufficient for
authorisation and implicitly, also for audit purposes, but is not what
many in the information security sector would view as sufficient for
integrity.

\section{Disk and file encryption}\label{sec:dfe}

\begin{userstories}
\USAdata
\end{userstories}

Disk and file encryption is purely about confidentiality, with some
large examples described by~\citeN{mSecurosis1}.  We suggest that this is the
simplest of our selection of problems. Essentially, media can be
lost: making it hard for the records on that media to be (ab)used
is an obligation in most data processing scenarios.

Examples are easy to find in the media; we are aware of local
cases, \eg,~involving sensitive medical records.  The Information
Commissioner's Office has a range of press
releases~\cite{mICO} detailing some of these incidents.  In
terms of risk assessment, we suggest that this is one of the most significant
risks facing most organisations.  Whereas we argue in our earlier
stories that the integrity of information is relatively rarely
challenged, there is a high likelihood of accidental loss of storage media and computing
equipment; similarly the type of
information can range from trivial to highly compromising.

We initially suggest that this should be relatively easy to manage.
A range of software is available, including paid-for and free applications.
Small installations can rely on simple use of passwords, while larger
organisations may use some form of enterprise management capabilities
such as Symantec's PGP Whole Disk Encryption.

As usual, we observe that the practicalities are not so easy.
Discussions with local SMEs during short (one-day) basic IT security courses demonstrate that some simply do not
recognise the need to protect data from inappropriate disclosure although the need for antivirus software is
commonly recognised.  Those that do sometimes suffer from choice-paralysis: how
do non-experts choose a suitable piece of software?  

Built-in options are little better: ``scary'' but otherwise correct dialogues
about encryption passwords being critical deter users.  The
well-known costs associated with managing additional software,
handling keys, issues with backup and recovery, \etc, become relevant.  As a
final remark, we note that a small, but significant minority of our
undergraduates found TrueCrypt's dialogues confusing: these students
managed to overwrite existing files when they were trying to create
new file containers.

However, given the risks for most users, we argue that any
reasonable disk encryption is effective, as the aim is to prevent
compromise due to accidental loss and casual thieves, albeit not effective
at dealing with determined attackers.

\section{Discussion}\label{sec:discussion}

We have examined a range of common applications.  We now examine four
overarching themes in relation to user-visible applications of cryptography:
\begin{itemize}
\item risk and value;
\item deployment problems; 
\item endpoint security; and
\item trust problems.
\end{itemize}

\subsection{Risk and value}\label{sec:riskValue}

We have identified a range of risks in our user stories above.  We can place them
into three groups:
\begin{itemize}
\item risks best mitigated by user-visible applications of cryptography;
\item low risks; and
\item risks that are mitigated by legal, societal or
  other technological measures.
\end{itemize}
We now take these in turn.

\subsubsection{Risks mitigated by user-visible applications of cryptography}

The \USTdata\ user story is an outlier compared to our other user
stories.  It demonstrates an effective mitigation of the risk using
user-visible applications of cryptography, although problems such as the
(mis)management of encryption keys can occur.  Essentially it can
convert accidental and inevitable loss of readable data on portable
media into the loss of encrypted data.  Thus in this case, the use of cryptography is valuable compared to the risk.  Even
then, it can be automated further by inclusion in the boot
process.  A diligent attempt to use encryption can form part of the
management of the legal risk from, say, the UK's Data Protection Act 1998.

\subsubsection{Low risks} 
The risks in some of these user stories are low as the impact of a
breach is low, for example in social websites and social email.  The lower this
risk, the less justification there is for the costs ---time, effort
and money--- for user-visible applications of cryptography as distinct from
technologies such as opportunistic encryption.  
Users perceiving a low impact, whether consciously or unconsciously, are unlikely to attempt to mitigate that risk.

\subsubsection{Risks mitigated by legal, societal or
  other technological measures}

The remaining medium or high risks can be mitigated by other means.  

Although we have concentrated on UK (albeit primarily English
  and Welsh and related systems such as Canadian and Australian) law, 
  the legislative situation is similar in other
  jurisdictions.  For example, for data protection issues, European
  countries have their own implementations of the 1995 European Union
  Data Protection Directive.  
Procedural and audit safeguards are often in place, particularly relevant for environments where a relatively large number of users may legitimately access data
(such as in the health and law enforcement sectors).

  Notably, cryptographic signatures typically have no added legal
  value over other types of signatures (as described in
  section~\ref{sec:legal-context}).  This applies particularly to the
  \USTwecommerce, \USTwbanking\ and \USTecontract\ user stories, and
  to a lesser degree, the two scenarios discussed in
  section~\ref{sec:formSigning}.  The mitigation in all these cases is
  that the parties have recourse to the legal systems, where courts
  would be asked to decide if a contract existed.  A simple email
  without a cryptographic signature may be sufficient for a court.  
  Chapter 8 of~\citeN{Mason12} discusses issues of liability further.

 For financial transactions, reactive monitoring systems, as exemplified by credit card companies,
  identify anomalous patterns of use which triggers out-of-band
  authorisation to the credit card holder.  This monitoring, along
  with legal guarantees limiting the risk to the card holder, can
  substantially reduce the risk at least to the card holder; the merchant may take on greater risk, along with the issuing bank.  However, we commented earlier on the variability of such legal protection.  
The advertised guarantees to account holders and the relatively low likelihood of any particular individual becoming a victim versus the obvious convenience of online banking can reasonably account for the popularity of online banking.  The issues with CAs and SSL simply do not pose a sufficient problem for these users
to decline to use online banking and similar services.

\subsubsection{Limitations to our evaluation of risks}

There are limitations to our evaluation of security risks.  
The argument advanced so far is qualitative.  A finer-grained analysis
requires quantitative data and suitable  objective metrics, as
suggested by~\citeN{Stolfo+11}.  In their work, they 
describe at least three kinds of  adversaries
\begin{itemize}
\item nation state actor,
\item expert operator adversary, and
\item insider expert developer.
\end{itemize}
However, evidence from reported incidents suggests that 
casual, opportunistic and accidental risks such
as phishing and inadvertently losing storage media
should be of greater concern to most end-users of the type we are
concerned with in this work.  Moreover, in the absence of
strong compartmentalisation, the technologies discussed in this paper are
unlikely to deter or restrain any of \citeANP{Stolfo+11}'s adversaries.

Moreover, the evaluation of value or impact is notoriously dependent
on the viewpoint of individual stakeholders.  For example, Schneier
refers to externalities~\cite{mSchneier07} and \citeN{Ackerman+99}
discuss the varying value of different types of information according
to individual preferences.

\subsection{Deployment problems}

The previous section has illustrated the value or lack thereof of
user-visible applications of cryptography for mitigating the risks in our user
stories.  We now examine how these mitigations are sometimes
undermined in practice.

There are two ways this undermining occurs:
\begin{itemize}
\item through lack of individual and organisational awareness for the
  need for cryptography; and
\item (mis)use of that cryptography.
\end{itemize}

The former point applies to both the selection and implementation of
suitable systems by organisations, as well as actual use by individual
end-users.  For example, organisations may not recognise the need for
encryption of sensitive data.  Even if an organisation does recognise
this need, individual end-users may not recognise it, or may disregard it for other
reasons.  Thus organisations promulgate their need via policies and
procedures.  Other authors have also commented on awareness, as well
as technological and regulatory issues.  For example,
\citeN{Srivastava09} considers the Australian environment and remarks
``there is significant evidence of Australian businesses' lack of
awareness and understanding of electronic signatures and the
associated legislation, despite a regulatory framework to facilitate
their use''.

The correlation between risk and the awareness of need for
cryptography is unclear.  This is illustrated in our user
stories.  For example, in the lowest risk user stories such as
social email, we would not expect any awareness of need.  At the
other end of the scale such as online banking, awareness should be
high in part due to media coverage.  Interaction with users suggests
this is the case.  However, other user stories are problematic.  The
risks with handling removable media are often not recognised, as
evidenced in the UK by reports from the Information Commissioner's
Office~\cite{mICO}.  Additionally, inadvisable software installation
is implicated in malware infections on end-user computers.  There is
little surprising in terms of awareness here, and it remains an open
question how end-users can be made aware.
Automation (discussed shortly) remains the most obvious option.

Mitigation of risk using cryptography is also undermined
by intentional or accidental misuse.  Of particular interest is the
usability of these tools.
We have previously asserted that  users of interest to us have basic
computer skills, \eg, word processing and email.  Requiring them to
directly operate cryptographic software poses a substantial problem when considering
usability~\cite{Whitten+05}.
The problems continue, and anecdotal evidence is in good supply.  For
example, Roger Grimes says
\begin{quote}
  ``Case in point: I routinely use Pretty Good Privacy (PGP) and SMIME
  to secure e-mails and file transfers. Yet frequently, even somewhat
  knowledgeable IT security people get confused about which keys to
  use when. In order to for someone to send me encrypted content, I
  need to send that person my public key. Similarly, I need the
  recipient's public key so that I can send him or her encrypted
  content. We should never share private keys. That's why they are
  called private. Pretty simple --- or so you would think. More often
  than not, if the person isn't overly familiar with PGP/SMIME, even
  if they've been using it, they send me their private key.

  ``Being the good citizen that I am, I delete their private key and
  ask again for their public key, explaining that with their private
  key, I could be them, for all digital purposes. About half the newly
  educated group then sends back my public key back or, if they're
  using PGP, their private key ring, which contains all their private
  keys. You might think that I'm making this stuff up, but it's pretty
  much been this way with PKI and PGP exchanges since they were
  invented. PGP's own Phil Zimmerman has often written on this
  subject.''~\cite{mGrimes09}
\end{quote}
Simplicity of use is obviously beneficial.  We might reasonably
consider that user-visible applications of cryptography have an inherent requirement
for user effort, and that deployment involves consideration of
training requirements.  However, our end-users are trying to achieve
relatively simple tasks; the computers are a means to an end and our
user might view the computer as nothing more than a tool like a
washing machine.  Thus our argument is that in the scenarios we
consider, asking for any significant user effort to understand and
correctly use these cryptographic features is unreasonable and likely
to result in non-conformance and inadvertent misuse.

\subsection{Endpoint security}\label{sec:endpoint}

The problem is not limited to the tools alone.  General issues of
security hygiene arise such as leaving computers unlocked in
vulnerable environments; indeed, some  organisations 
such as  universities
and Internet cafes
disable the screen locks\label{pg:screenlocks} to prevent
monopolisation of shared computers.  Poor password practice is
common \cite{mPassword08}; a typical scenario is demonstrated
by technicians as illustrated by ``Ted'', a technician
who had arrived to update some software for user ``Alice'' who had
gone to speak to a colleague on the other side of their open-plan
workspace.

\begin{centering}
  \begin{tabular}{ll}
    Ted (shouting across the room)&“Alice, what's your password?”\\
    Alice (shouting back)&“It's (her real password).”
  \end{tabular}

\end{centering}
\noindent This particular workplace dealt with medically-sensitive information,
and all the workers were (at least in theory) aware of the need to
control this information.
This same workplace, despite using a relatively modern mail server,
insisted that the \emph{only} way for one user to access the email of
another user while that user was on long-term leave was to ask the
absent user for her password.  The more appropriate mechanism
involved auditable delegation via the mail server.

Thus the endpoints, the computers our end-users are using, are a
significant weak link.   Data is accessible on these machines, and
some store CMS keys with no further protection.  Gene Spafford is quoted as saying
\begin{quote}
  ``Using encryption on the Internet is the equivalent of arranging
   an armored car to deliver credit-card information from someone
   living in a cardboard box to someone living on a park bench.''
\end{quote}

The changing consumer computer environment produces challenges: desktop machines are relatively well-understood, but the use of mobile devices with operating systems such as iOS and Android pose additional challenges.  They are easier to steal, and in some cases, have a more limited access control model~\cite{Hayashi+12}.

Some attempts to secure endpoints address the difficulty of handling many, good quality passwords.  
For example, IDSpace proposes a single user interface that supports a range of existing identity management technologies~\cite{Alsinani+12}.
More ambitious is Stajano's Pico~\cite{Stajano11}.  This involves a proposed hardware device that takes over the role of authentication.  Of course, this brings issues of authenticating the user to the Pico; the use of a swarm of picosiblings (using $k$ out of $n$ secret sharing) and biometrics is suggested.  A very broad discussion of proposals for replacing passwords is due to \citeN{Bonneau+12}.

Lastly, reidentification and recovery issues provide a potential weak point in many deployed systems.  Forcing password recovery via some means an attacker controls is well known.

The problem of endpoint security relates back to issues of liability, and the legal and contractual context of these transactions.
ENISA states that
\begin{quote}
  ``many online banking systems dangerously rely on PCs being secure,
  but banks should instead presume all customer PCs are
  infected.''~\cite{mENISA}
\end{quote}
On the same theme, Krebs comments
\begin{quote}
  ``No online banking authentication system works unless it starts
  with the premise that the customer’s machine is already compromised
  by malware that gives thieves complete control over the customer
  system. But for better or worse, the commercial banks have no
  (dis)incentive to do much to improve the integrity of online banking
  transactions because the current regulations effectively hold them
  blameless when a customer loses money.''~\cite{mKrebs}
\end{quote}
The latter point is a significant point of variation.  Recovery of funds lost to criminal activity vary amongst different jurisdictions.  Even where regulations apparently are in the consumers' favour, the reality can be different~\cite{mFIPR,mMason2}.

\subsection{Trust problems}\label{sec:trustRoots}

So far, we have discussed the value of user-visible applications of cryptography for
the mitigation of risks in our user stories.  We now take a more
holistic view and examine trust.

We place trust in cryptographic protocols which are believed to be
sound, and their implementation due to testing, review, credibility,
\etc\  But for most of our user stories, data confidentiality being
an exception, these depend on trusted third parties (TTPs): the certificate authorities.  These
TTPs are used to bootstrap trust when there is no prior relationship
between the first and second parties.
The users trust the CA to check the identity of the service provider and correctly link the offered X.509 certificate to that identity.

However, we have seen that this trust may be misplaced: in
section~\ref{sec:web} we noted issues with bootstrapping trust
relationships~\cite{Perlman99} and highlighted the Comodo and more recent
compromises~\cite{mComodo1}.  In section~\ref{sec:email}, we remarked
that some of these issues can be mitigated by the local network
infrastructure, \eg, to allow opportunistic encryption, if that local infrastructure  is
trusted.

There are alternatives to a  naming scheme that is world-wide (\ie,
requesting a certificate from a well-known CA); one  is to
use local, closed CAs.  A further option is  \emph{web of trust}
style keying.  These have been discussed at some length earlier.  

Others attempt to fix the existing CA environment include “pinning” which
whitelists public keys that are expected by a particular browser to
make it harder for untrustworthy certificate chains to go undetected.
More interesting is the use of multiple notaries as illustrated in the \textit{Perspectives} project~\cite{Wendlandt+08} and the subsequent Convergence add-on/daemon~\cite{Marlinspike12}: both have users selecting notaries that they trust rather than relying on the default root CAs provided in (say) a web browser.  
However our earlier arguments suggest that casual users will not be willing to engage in any additional work to choose their trust relationships as there is no real improvement in their situation given the user effort required; this broadly matches with Herley's conclusions~\cite{Herley09}.  
A broader discussion about how trust operates in societies is given by \citeN{Schneier12}.

Despite the issues with CAs described earlier, we may ask why companies such as Verisign and Entrust amongst others can run a business selling SSL certificates. 
We suggest that there are two major factors:
\begin{itemize}
\item regulatory compliance, such as the PCI SSC Data Security Standard~\cite{PCIDSS}; and
\item the inclusion of their root certificates in major web browser installation packages.
\end{itemize}
For the relatively low cost per unit, an individual business will not need to consider the purchase for long; yet the vendors have a wide range of potential buyers and this is a business that scales well.

In practice, the TTP infrastructure might not be trustworthy but users
use it  anyway, and when warning dialogues appear they are
disregarded.  Some protocols, \eg, SSH, can record the known host keys
and warn when it changes in the style of key continuity management (discussed earlier in section~\ref{sec:gateways}); similarly, we observe that many users
disregard this and continue with their connection.  These points relate to
the awareness and education issues previously highlighted.

\section{Conclusions}\label{sec:conclusions}

We have examined user-visible applications of cryptography.  In part, our analysis has been structured by ten user stories in the context of the UK
regulatory environment.  None of these are what would be classically
considered ``critical systems''.  Instead, they are routine,
day-to-day scenarios.  These user stories have been addressed by scenario-based analysis and we have
examined the balance of risk and value of user-visible applications of cryptography,
and the subsidiary deployment and trust issues.

We see that despite the apparent problems, particularly those associated with deployment, endpoint security and trust (especially of CAs), the deployed systems work relatively well.  
Our survey suggests that this is due to the presence of mitigating factors, such as guarantees to bank account holders and recourse to the legal system.

We return to the three categories of user-visible applications of cryptography from section~\ref{sec:categories}.
\begin{enumerate}
\item Direct, elective invocation of a cryptographic tool is very rare in the user populations.
\item  Indirect but still explicit, elective use of cryptography is valuable in the \USTdata\ user story, but does not appear elsewhere.
\item Implicit, background use of cryptography accounts for most of the usage.  Even where problems occur, we argue that users ignore or otherwise accept the risks (similar to \citeN{Herley09}'s argument) or are content for other mitigations to operate.
\end{enumerate}

Any application of user-visible cryptography must make sense in that particular context. 
Our survey illustrates that the  social/legal
framework often does not demand any cryptographic mechanism, or that users, software and/or the
infrastructure compromise its effectiveness.  A significant exception
is the use of encryption to ensure data confidentiality, particularly
for removable media.  
In general, the potential security issues are essentially
peripheral to the user's concerns; the users are trying to achieve some
other objective using the computer as a tool.

In the work leading to this survey, we see three particularly relevant areas for further work:
\begin{itemize}
\item metrics to objectively establish quantitative measures for the value of user-visible applications of cryptography in these types of user story;
\item usability and education issues, as discussed by other authors in section~\ref{sec:user}; and
\item the balance of automation and control, which we discuss next.
\end{itemize}

We have remarked that the endpoint computing devices are a significant
vulnerability.  We suggest the following definitions:
\begin{description}
\item[Automation] This relates to the computer making decisions with
  minimal, if any, user intervention, and incorporates elements such
  as robustness.  For example, if the user is asked to handle a failed
  verification that could be due to network problems,
  incompatibilities of cryptography, inadvertently modified files or
  malicious attack, this is a lack of automation.
\item[Control] This is the ability of a principal to dictate the usage
  of a computing device, access to information, and may include some
  authority or opinion over trust models and trust roots.  

  A locked-down computer, where the end-user cannot install anything
  that is not approved by the original vendor could be viewed as
  overly paternalistic, but could, if the vendor has suitable
  judgement, increase the possible automation in terms of certificate
  authorities.  These are more computer appliances  than general
  purpose computers.  Of course, this approach is anathema to the
  free/libre open source software (FLOSS) community.
\end{description}
We remark firstly that automation and control are not necessarily
opposing, although there is an obvious tension.  Moreover, they are
potentially different for each stakeholder, \eg, computer user \vs\
system administration \vs\ software publisher \vs\ software developer.
Both automation and control need to be balanced against the overall risk.

Earlier comments about usability lead us to conclude that one partial
mitigation is for software to require proactive changes of security
settings rather than reactive changes.  As an example, consider access
to a website using an SSL certificate that has not been suitably
signed.  A reactive approach allows the user to add exceptions.
Instead, the software could simply refuse access as an extreme level
of automation.  The proactive approach requires that the user take
deliberate action unprompted by access to the web server: this gives
the user some control, but with reduced compromise of automation.
Essentially, this is to make it harder for users to say ``I don't
care, just let me access the web site''.  A multitude of controls and
fine-grained options has no value if the user will click ``okay, get
on with it'' no matter what; perhaps heavy automation with strong
controls is a suitable amelioration.  This deserves further
user-focused study, perhaps by systematic repetition of our scenarios but examining the interactions of particular interfaces more closely.  A practical implication is that simplified,
constrained applications may be a short-term compromise that reduces
the risks
while still allowing sufficient utility.

\section*{Acknowledgements}

This paper has been through several revisions over the past two or so
years.  We thank the various anonymous reviewers who have made
valuable comments on this work.

\bibliographystyle{acmtrans}
\bibliography{refsauto,refsman}

\end{document}